# Preoperative Decline and Postoperative Recovery of Wearable-Derived Physical Activity Over a Four-Year Perioperative Period in Total Knee and Hip Arthroplasty: Evidence from the All of Us Research Program

Yuezhou Zhang[1], PhD; Amos Folarin[1,2,3,4,5], PhD; Callum Stewart[1], PhD; Hyunju Kim[1], PhD; Rongrong Zhong[1,6], MD; Shaoxiong Sun[1,7], PhD; Richard JB Dobson[1,2,3,4,5], PhD

[1]Department of Biostatistics & Health Informatics, Institute of Psychiatry, Psychology and Neuroscience, King's College London, London, United Kingdom

[2]Institute of Health Informatics, University College London, London, United Kingdom

[3]NIHR Biomedical Research Centre at South London and Maudsley NHS Foundation Trust, London, United Kingdom

[4]NIHR Biomedical Research Centre at University College London Hospitals NHS Foundation Trust, London, United Kingdom

[5]Health Data Research UK, University College London, London, United Kingdom

[6]Clinical Research Center & Division of Mood Disorders, Shanghai Mental Health Center, Shanghai Jiao Tong University School of Medicine, Shanghai, China

[7]Department of Computer Science, University of Sheffield, Sheffield, United Kingdom

**Abstract**

Total knee arthroplasty (TKA) and total hip arthroplasty (THA) improve symptoms in end-stage osteoarthritis, yet long-term objective characterization of perioperative physical activity trajectories remains limited. We conducted a longitudinal observational study within the All of Us Research Program dataset, linking electronic health records with continuous Fitbit-derived step count data over a four-year perioperative window (two years before and two years after arthroplasty). Piecewise linear mixed-effects models characterized preoperative declines and postoperative recovery trajectories, and time-to-recovery was evaluated using Kaplan–Meier curves and Cox proportional hazards models under remote and immediate preoperative physical activity baseline definitions. Among 238 participants (147 TKA; 91 THA), both procedures exhibited progressive preoperative decline with distinct procedure-specific patterns and staged postoperative recovery: rapid improvement during weeks 1–6, decelerating gains through weeks 7–19/20, and subsequent stabilization through week 104. Recovery to remote and immediate baselines differed in timing (median 22 vs 13 weeks) and associated predictors. Higher immediate preoperative activity was associated with greater likelihood of recovery to habitual activity levels, underscoring the relevance of preoperative functional reserve and surgical timing. These findings demonstrate the potential of long-term wearable monitoring to refine assessment of functional outcomes, guide recovery expectations, and support perioperative management.

**Introduction**

Osteoarthritis is a prevalent and disabling condition affecting more than 500 million people worldwide and imposing a broad burden on individuals, including chronic pain, functional impairment, activity limitation, and reduced quality of life [1]. Total knee arthroplasty (TKA) and total hip arthroplasty (THA) are among the most frequently performed and increasingly utilized orthopaedic procedures worldwide and are highly effective in relieving pain and improving function in end-stage osteoarthritis [2]. Despite their overall effectiveness, a considerable proportion of patients report persistent pain [3, 4], limited improvements in physical activity or functional performance [5-7], or dissatisfaction [8] in the months to years after arthroplasty. Emerging evidence further suggests that preoperative physical activity levels are associated with postoperative outcomes after TKA and THA [9, 10]. Together, these findings underscore the need to characterize preoperative and postoperative physical activity trajectories and to identify factors that influence postoperative recovery.

Traditional assessments of preoperative status and recovery following TKA and THA primarily rely on patient-reported outcome measures (PROMs) and intermittent clinical evaluations [11-13]. Although clinically informative, these approaches are inherently subjective and susceptible to recall bias, time-intensive with often limited patient compliance, and inadequate for characterizing dynamic physical activity recovery trajectories over time [14-16]. Moreover, the accuracy of patient recall of preoperative symptoms declines substantially beyond three months postoperatively, highlighting challenges in reliably assessing baseline functional status [17].

Advances in sensor technology and the widespread adoption of consumer wearable devices have enabled cost-effective, low-burden, and objective monitoring of real-world physical activity at scale [18-20]. For example, wristband devices such as Fitbit trackers can capture daily step count—a simple and intuitive objective measure of overall activity volume [21]— over extended periods with minimal user burden [22]. Leveraging these capabilities, a growing number of studies have used wearable devices to quantify postoperative recovery in physical activity following TKA and THA, demonstrating feasibility and establishing objective benchmarks of functional recovery in real-world settings [23-28]. However, several limitations remain. First, many studies summarize activity changes at prespecified postoperative time points (e.g., 1, 3, and 6 months) or present descriptive visualizations rather than explicitly modeling recovery trajectories. Second, preoperative wearable data, when available, typically cover only a short period before surgery (often less than one month), and recovery is often evaluated relative to immediate preoperative physical activity levels, which may reflect disease-related decline rather than longer-term habitual baseline function. Third, postoperative follow-up is often limited to early and mid-term recovery (weeks to 6–12 months), with less characterization of longer-term recovery trajectories and stabilization.

In addition, TKA and THA may involve distinct biomechanical and rehabilitation processes, reflecting differences in joint mechanics, muscle function, and recovery pathways [29-31]. Wearable-based studies have also demonstrated differences in postoperative recovery trajectories between TKA and THA [32], and some studies suggest that recovery in physical activity and mobility may occur more rapidly following THA than TKA [23, 27, 33]. However, these studies typically rely on comparisons at prespecified postoperative time points and define recovery relative to immediate preoperative activity levels, with limited characterization of preoperative trajectories or comparison of declines in physical activity before surgery between procedures.

To address these gaps, longer-term continuous wearable-derived physical activity data before and after arthroplasty are needed. The All of Us Research Program (AoURP) provides a uniquely integrated resource linking longitudinal EHRs with historical Fitbit-derived activity data, enabling identification of arthroplasty timing and retrospective characterization of physical activity trajectories over extended pre- and postoperative periods [34]. Here, we aimed to investigate whether functional decline and recovery surrounding arthroplasty exhibit quantifiable, staged, and predictable patterns in wearable-derived physical activity. Specifically, we first characterized preoperative and postoperative step count trajectories over the two years before and two years after surgery for TKA and THA using piecewise linear mixed-effects models to identify distinct phases of functional decline and recovery. Second, we defined immediate and remote preoperative physical activity baselines and applied survival analysis to evaluate time to recovery and identify socio-demographic and baseline activity-related determinants of recovery.

**Materials and Methods**

**Data Source and Study Population**

We conducted a longitudinal observational cohort study within the AoURP, an ongoing national longitudinal cohort funded by the US National Institutes of Health with the goal of enrolling at least one million participants [34]. The overall study design and data collection procedures have been described previously [34, 35].

This analysis used the controlled tier dataset (version 8; C2024Q3R8), including participants enrolled between May 2017 and October 2023. Socio-demographic characteristics and baseline information were collected at digital enrollment. For participants who consented to share EHR and Fitbit data, historical (pre-enrollment) EHR records and Fitbit data were made available through participating healthcare provider organizations and linked Google Fitbit accounts, respectively [34-36]. In this study, deidentified data from the AoURP were accessed exclusively by authorized authors who completed the mandated Responsible Conduct of Research training, with all analyses performed within the secure Researcher Workbench platform.

**Identification of Arthroplasty Procedures**

TKA and THA procedures and their corresponding dates were retrospectively identified from EHR data within the AoURP controlled tier dataset. Procedures were ascertained using Current Procedural Terminology (CPT-4) codes, including CPT4-27447 for TKA and CPT4-27130 for THA. Eligible procedures were restricted to those occurring during periods with available Fitbit data. For participants with multiple eligible arthroplasties during the observation window, only the first procedure was included.

**Wearable Data Processing and Temporal Alignment**

Daily step count data were obtained from linked Fitbit devices within the AoURP controlled tier dataset. Step counts were temporally aligned to the procedure date, defined as day 0, and aggregated into consecutive, non-overlapping 7-day intervals to derive weekly-averaged daily step counts. Preoperative time was indexed from weeks −104 to −1, and postoperative time from weeks 1 to 104, with postoperative week 1 corresponding to days 0–6. To ensure data reliability and mitigate bias from sparse observations, weeks were retained only if they contained at least three observed days; weeks with insufficient data coverage were excluded

**Modeling of Preoperative Physical Activity Trajectories**

To characterize deterioration in physical activity prior to surgery, piecewise linear mixed-effects models [37] with participant-specific random intercepts were fitted separately for TKA and THA. A retrospective time scale spanning weeks −104 to −1 was used to represent the two-year preoperative period, with week −1 corresponding to the week immediately preceding surgery. Participants were required to have at least 12 valid weeks of wearable-derived step count data during this period. Relative week was modeled as a continuous variable, and weekly-averaged daily step count was treated as the outcome. One knot (change point) was specified to allow estimation of distinct preoperative phases, and its optimal location was determined using a grid search across candidate weeks, with model selection based on the lowest Akaike Information Criterion (AIC). Models were adjusted for age, sex, and body mass index (BMI), as these factors are well-established determinants of physical activity levels and may confound the association between time relative to surgery and activity trajectories [38, 39].

**Modeling of Postoperative Recovery Trajectories**

Postoperative recovery was hypothesized to follow a multi-phase trajectory based on prior literature [24, 25, 40, 41] and visual inspection of the data. Physical activity during weeks 1–104 after surgery was therefore modeled using piecewise linear mixed-effects regression with participant-specific random intercepts, fitted separately for TKA and THA. Two knots were specified to allow three distinct recovery phases, and their optimal locations were identified using a grid search across candidate postoperative weeks, with model selection based on the lowest AIC. Participants were required to have at least 12 valid postoperative weeks of wearable-derived step count data. Within each segment, relative postoperative week was treated as a continuous variable. Weekly-averaged daily step count was modeled as the outcome with adjustment for age, sex, and BMI.

**Definition of Recovery in Physical Activity and Time-to-Recovery Analysis**

Recovery in physical activity was defined as the first restoration of weekly-averaged daily step count to or above a predefined preoperative baseline threshold. To minimize misclassification due to short-term fluctuations, recovery required two consecutive weeks meeting or exceeding the threshold.

Two preoperative physical activity baselines were applied:

1) Remote preoperative baseline: average daily step count during weeks −104 to −55, representing longer-term habitual activity;
2) Immediate preoperative baseline: average daily step count during weeks −4 to −1, representing activity immediately preceding surgery.

Participants were required to have at least one valid week within the corresponding baseline window and at least 12 valid postoperative weeks to be included in the time-to-recovery analysis.

Time-to-recovery was evaluated using Kaplan–Meier curves [42]. Cox proportional hazards (CoxPH) models [43] were fitted to estimate hazard ratios (HRs) for associations between preoperative predictors and time-to-recovery. An HR > 1 indicated a higher likelihood of achieving recovery. Preoperative predictors included sociodemographic characteristics (age, sex, marital status, employment status, education, income), BMI, surgery type (TKA vs THA), and baseline activity levels. Separate CoxPH models were fitted for recovery defined using remote and immediate baseline thresholds. The proportional hazards assumption was assessed using scaled Schoenfeld residuals [43].

**Results**

**Participant Characteristics**

A total of 238 participants were included in at least one analytic cohort, comprising 147 TKA and 91 THA procedures. Sample sizes varied slightly across analyses (preoperative trajectory modeling, postoperative trajectory modeling, and time-to-recovery analysis) according to data availability and eligibility criteria. The mean age at consent was 64.9 years (SD 8.3), 71.8% were female, and mean BMI was 30.9 kg/m² (SD 6.6). Most participants had completed a college degree (71.4%), and 37.0% were employed at baseline. Socio-demographic characteristics were comparable between TKA and THA groups (all P > 0.05; Table 1). Figure 1A displays the observed weekly-averaged daily step counts relative to the procedure date, from 104 weeks before to 104 weeks after arthroplasty.

**Preoperative Physical Activity Trajectories**

Piecewise linear mixed-effects models revealed distinct patterns of preoperative physical activity decline in TKA and THA (Table 2; Figure 1B). At week −104, weekly-averaged daily step counts were similar between groups (TKA 7,510.1 [95% CI 6,953.3–8,066.8] vs THA 7,654.4 [6,931.7–8,377.1] steps). During the early preoperative period (week -104 to week -55), both groups showed only modest declines in weekly-averaged daily step counts (TKA −7.2 steps/week; THA −4.8 steps/week; both P < 0.001), after which their trajectories diverged. In TKA, weekly-averaged daily step count began to decline more rapidly 55 weeks before surgery (−28.8 steps/week; P < 0.001), whereas THA maintained a gradual decline until 16 weeks preoperatively, followed by a sharp reduction in the final months before surgery (−58.5 steps/week; P < 0.001). By the week immediately preceding surgery (week -1), weekly-averaged daily step counts had decreased to 5,989.9 [5,440.0–6,539.9] in TKA and 6,425.0 [5,685.0–7,165.0] in THA.

**Postoperative Recovery Trajectories**

Three-phase piecewise linear mixed-effects models identified staged recovery patterns in both procedures (Table 3; Figure 1C), with optimal knots at weeks 6 and 20 in TKA and weeks 6 and 19 in THA. In the early postoperative period (weeks 1–6), weekly-averaged daily step counts increased rapidly from 1,438.8 [823.8–2,053.7] at week 1 to 4,781.2 [4,220.1–5,342.2] at week 6 in TKA, and from 1,111.2 [339.4–1,883.0] to 5,962.8 [5,247.6–6,678.0] in THA, with a steeper weekly slope in THA than TKA(+970.3 vs +668.5 steps/week; both P < 0.001). During the intermediate phase (weeks 7–20 in TKA; 7–19 in THA), the rate of improvement slowed substantially (TKA +111.6 steps/week; THA +89.8 steps/week; both P < 0.001), with step counts reaching 6,343.7 [5,801.4–6,886.1] at week 20 in TKA and 7,130.1 [6,438.3–7,822.0] at week 19 in THA. Thereafter, trajectories plateaued through week 104, with minimal weekly changes (TKA +3.4; THA −1.4 steps/week), stabilizing at 6,628.7 [6,082.2–7,175.1] in TKA and 7,014.2 [6,315.4–7,713.1] in THA.

**Time-to-Recovery and Associated Factors**

Using the remote preoperative baseline to define recovery in physical activity (see Materials and Methods), 152 participants were eligible (TKA 95; THA 57), and 73.7% achieved recovery within 104 weeks. Median time to recovery was 22 [95% CI 20–29] weeks overall (TKA 25 [20–33] weeks; THA 21[13–36] weeks), with no significant difference between procedures.

In terms of immediate preoperative baseline, 185 participants were eligible, and 85.4% achieved recovery. Median time to recovery was 13 [11–15] weeks overall (TKA 16 [13–21] weeks; THA 9 [8–12] weeks), with significantly faster recovery in THA (log-rank P = 0.001). In accordance with AoURP reporting policies, cell sizes fewer than 20 cannot be disclosed; therefore, procedure-specific counts of non-recovery or censoring are not reported separately.

In the CoxPH model for recovery to the remote preoperative baseline (Table 4), higher immediate preoperative physical activity was associated with faster recovery (HR 1.49 per 1000 steps; P < 0.001), whereas higher remote baseline activity was associated with a lower probability of recovery (HR 0.69; P < 0.001). Older age was independently associated with slower recovery (HR 0.96 per year; P = 0.007). In terms of CoxPH model for the immediate preoperative baseline, surgery type was the only significant predictor, with TKA associated with lower likelihood of recovery compared with THA (HR 0.55; P = 0.001; Supplementary Table 1).

**Discussion**

In this retrospective analysis of participants within the AoURP who underwent TKA or THA and had linked wearable data available, we quantified perioperative functional trajectories over a four-year period using integrated EHR and continuous step count data. We demonstrate three key findings. First, physical activity declined progressively prior to surgery, with distinct procedure-specific temporal patterns. Second, postoperative recovery followed a consistent staged trajectory characterized by rapid early improvement, slower intermediate gains, and long-term stabilization. Third, lower immediate preoperative physical activity levels and older age were associated with a lower likelihood of recovery to long-term habitual activity levels. In addition, recovery defined relative to remote versus immediate preoperative baselines differed in both time course and associated determinants, highlighting the importance of baseline definition when evaluating postoperative recovery in physical activity. Together, these findings highlight the value of continuous wearable data for objectively quantifying perioperative physical activity trajectories, informing recovery expectations, and identifying factors that may guide surgical timing for better recovery.

Our analyses extend prior work by quantifying long-term preoperative physical activity trajectories. While both TKA and THA exhibited progressive declines, TKA showed a prolonged, sustained decrease, whereas THA remained relatively stable until a steeper decline emerged in the final months preceding surgery. These distinct trajectories may reflect known differences in joint biomechanics, symptom progression, and functional impairment between knee and hip osteoarthritis [44]. Moreover, some studies further suggest that patients with hip osteoarthritis may undergo joint replacement earlier in the disease course than those with knee osteoarthritis [45], which may partially explain the later onset of accelerated decline observed in THA. Although these procedure-specific preoperative patterns require further validation, our findings underscore the value of extended preoperative monitoring and highlight heterogeneity in decline trajectories preceding arthroplasty.

Postoperative recovery in both procedures followed a consistent staged pattern identified using data-driven piecewise modeling. An initial rapid recovery phase during the first six weeks was followed by decelerating gains up to approximately 19–20 weeks and subsequent long-term stabilization. This staged recovery pattern aligns with prior studies reporting early improvement and later plateauing after arthroplasty [24, 25, 40, 41], while extending existing knowledge through continuous wearable-based measurement rather than prespecified postoperative time points. Notably, the early recovery slope was steeper in THA than TKA, consistent with previous reports of faster early functional improvement following hip arthroplasty [23, 29]. Although the exact timing of these recovery phases may vary across populations and study designs, our approach providing an objective and high-resolution characterization of recovery dynamics in real-world settings.

An important methodological and clinical insight from this study is the differentiation between physical activity recovery relative to an immediate preoperative baseline and a remote baseline representing longer-term habitual activity levels. We found that recovery to the immediate preoperative baseline occurred substantially earlier than recovery to the remote baseline, with a median difference of approximately nine weeks. This suggests that physical activity levels immediately prior to surgery may reflect a period of disease-related functional decline rather than typical functional capacity, consistent with the progressive nature of osteoarthritis [46] and our observed preoperative trajectories. Consequently, relying solely on the immediate preoperative baseline may overestimate both the speed and magnitude of recovery, whereas a remote baseline provides a more representative benchmark of habitual functional status and may offer a more clinically meaningful measure of recovery.

Our survival analysis showed that higher immediate preoperative activity levels were associated with a greater likelihood of recovery to habitual functional status (remote preoperative baseline), suggesting that better preserved functional capacity prior to surgery is associated with more favorable recovery outcomes. This finding aligns with prior evidence that preoperative functional status predicts postoperative recovery [9, 10, 47] and may inform surgical timing decisions. Older age was also associated with a lower probability of recovery, consistent with previous longitudinal studies of arthroplasty outcomes [48-50]. Additionally, as expected, higher remote baseline activity was associated with a lower probability of recovery within follow-up, likely reflecting a higher recovery threshold.

This study has several limitations. First, the bring-your-own-device design of wearable data collection in AoURP may introduce selection bias toward more health-conscious or technologically engaged individuals. Second, although the retrospective design enabled extended preoperative data, complementary clinical measures such as pain scores and PROMs were not consistently available, limiting multidimensional assessment of recovery. Third, although we used a remote preoperative baseline for capturing habitual activity levels, osteoarthritis-related functional decline may begin several years before surgery [51]. Future studies with longer-duration wearable data may further explore earlier stages of functional decline. Finally, despite the scale of AoURP, the subset of participants with continuous wearable data and confirmed arthroplasty remained relatively modest, warranting future validation in larger and more diverse cohorts.

**Conclusions**

Using linked EHR and four-year continuous wearable-derived activity data from the AoURP, we demonstrated that perioperative physical activity surrounding total knee and hip arthroplasty follows quantifiable, procedure-specific preoperative decline and staged postoperative recovery patterns. Recovery to remote and immediate preoperative baselines differed in both timing and associated factors, underscoring the importance of baseline definition in evaluating postoperative recovery in physical activity. Higher preoperative activity levels were associated with a greater likelihood of recovery, highlighting the relevance of functional reserve and surgical timing. Collectively, these findings demonstrate the potential of wearable monitoring to refine assessment of functional outcomes, guide recovery expectations, and support perioperative management.

**Acknowledgments**

We gratefully acknowledge All of Us participants for their contributions, without whom this research would not have been possible. We also thank the National Institutes of Health's All of Us Research Program for making available the participant data examined in this study.

**Funding**

Richard JB Dobson is supported by the following: (1) National Institute for Health and Care Research (NIHR) Biomedical Research Centre (BRC) at South London and Maudsley National Health Service (NHS) Foundation Trust and King's College London; (2) Health Data Research UK, which is funded by the UK Medical Research Council (MRC), Engineering and Physical Sciences Research Council, Economic and Social Research Council, Department of Health and Social Care (England), Chief Scientist Office of the Scottish Government Health and Social Care Directorates, Health and Social Care Research and Development Division (Welsh Government), Public Health Agency (Northern Ireland), British Heart Foundation, and Wellcome Trust; (3) the BigData@Heart Consortium, funded by the Innovative Medicines Initiative 2 Joint Undertaking (which receives support from the EU's Horizon 2020 research and innovation programme and European Federation of Pharmaceutical Industries and Associations [EFPIA], partnering with 20 academic and industry partners and European Society of Cardiology); (4) the NIHR University College London Hospitals BRC; (5) the NIHR BRC at South London and Maudsley (related to attendance at the American Medical Informatics Association) NHS Foundation Trust and King's College London; (6) the UK Research and Innovation (UKRI) London Medical Imaging & Artificial Intelligence Centre for Value Based Healthcare (AI4VBH); (7) the NIHR Applied Research Collaboration (ARC) South London at King's College Hospital NHS Foundation Trust; and (8) Wellcome Trust. The funders had no role in the design and conduct of the study; collection, management, analysis, and interpretation of the data; preparation, review, or approval of the manuscript; and decision to submit the manuscript for publication.

**Conflicts of Interest**

Richard JB Dobson and Amos Folarin are cofounders of Onsentia. Amos Folarin holds shares of Google.

**Data Availability**

To ensure participant privacy, the data used in this study are available to approved researchers through the All of Us Research Workbench (https://workbench.researchallofus.org/login) following registration, completion of required ethics training, and attestation of a data use agreement.

**Table 1. Sociodemographic characteristics of participants.**

| Characteristic | Overall (N=238) | Hip (n=91) | Knee (n=147) | P value |
|---|---|---|---|---|
| **Age, mean (SD)** | 64.9 (8.3) | 64.0 (9.0) | 65.5 (7.8) | 0.19 |
| **BMI, mean (SD)** | 30.9 (6.6) | 29.9 (6.4) | 31.5 (6.7) | 0.07 |
| **Sex, n (%)** | | | | 0.25 |
|     Female | 171 (71.8) | 61 (67.0) | 110 (74.8) | |
|     Male | 67 (28.2) | 30 (33.0) | 37 (25.2) | |
| **Education, n (%)** | | | | 0.346 |
|     College degree | 170 (71.4) | 67 (73.6) | 103 (70.1) | |
|     Some college/no college | 68 (28.6) | 24 (26.4) | 44 (29.9) | |
| **Income, n (%)** | | | | 0.06 |
|     High | 50 (21.0) | 26 (28.6) | 24 (16.3) | |
|     Mid/Low/Skip | 188 (79.0) | 65 (71.4) | 123 (83.7) | |
| **Employment, n (%)** | | | | 0.18 |
|     Working | 88 (37.0) | 39 (42.9) | 49 (33.3) | |
|     Not working | 150 (63.0) | 52 (57.1) | 98 (66.7) | |
| **Marital status, n (%)** | | | | >0.99 |
|     Married/Living with partner | 159 (66.8) | 61 (67.0) | 98 (66.7) | |
|     Never married/Divorced/Widowed/Separated | 79 (33.2) | 30 (33.0) | 49 (33.3) | |

TKA, total knee arthroplasty; THA, total hip arthroplasty; BMI, body mass index.

Education was categorized into three groups: college degree (college graduate or advanced degree), some college (college one to three), and no college (high school/GED or less). Income was grouped into three levels based on reported household income: high (≥$150,000), mid ($50,000–$150,000), and low/skip (<$50,000, prefer not to answer, or skip). Employment was categorized as working (employed for wages or self-employed) and not working (retired, homemaker, unable to work, or unemployed). Marital status was categorized as married/living with partner and never married/divorced/widowed/separated.

In accordance with All of Us Research Program reporting policies, categories with fewer than 20 participants in any cell cannot be reported to protect participant confidentiality; therefore, some categories were combined, and race and ethnicity information is not displayed.

**Table 2. Preoperative piecewise weekly change in daily step count before arthroplasty.**

| Group | Phase | Weekly change in daily step count, β (steps/week) | 95% CI | P value |
|---|---|---|---|---|
| TKA (N = 115) | Gradual decline phase (weeks −104 to −55) | −7.2 | −11.0 to −3.4 | <0.001 |
|  | Accelerated decline phase (weeks −54 to −1) | −28.8 | −32.1 to −25.6 | <0.001 |
| THA (N = 75) | Gradual decline phase (weeks −104 to −16) | −4.8 | −7.0 to −2.6 | <0.001 |
|  | Accelerated decline phase (weeks −15 to −1) | −58.5 | −74.6 to −42.4 | <0.001 |

TKA, total knee arthroplasty; THA, total hip arthroplasty.

Time was indexed relative to the procedure date and spanned the preoperative period from week −104 to week −1. Slopes (β) were estimated using two-phase piecewise linear mixed-effects models with participant-specific random intercepts. Knot locations were selected using the Akaike Information Criterion (AIC).

**Table 3. Postoperative piecewise weekly change in daily step count after arthroplasty.**

| Group | Phase | Weekly change in daily step count, β (steps/week) | 95% CI | P value |
|---|---|---|---|---|
| TKA (N = 120) | Rapid recovery phase (weeks 1–6) | 668.5 | 593.2 to 743.8 | <0.001 |
| | Decelerating recovery phase (weeks 7–20) | 111.6 | 96.9 to 126.4 | <0.001 |
| | Plateau phase (weeks 21–104) | 3.4 | 1.4 to 5.4 | 0.001 |
| THA (N = 81) | Rapid recovery phase (weeks 1–6) | 970.3 | 880.5 to 1,060.2 | <0.001 |
| | Decelerating recovery phase (weeks 7–19) | 89.8 | 70.8 to 108.8 | <0.001 |
| | Plateau phase (weeks 20–104) | −1.4 | −3.8 to 1.1 | 0.27 |

TKA, total knee arthroplasty; THA, total hip arthroplasty.

Time was indexed relative to the procedure date and spanned the postoperative period from week 1 to week 104 (week 1 corresponds to days 0–6 postoperatively). Slopes (β) were estimated using three-phase piecewise linear mixed-effects models with participant-specific random intercepts. Knot locations were selected using an Akaike Information Criterion (AIC)–based grid search.

**Table 4. Multivariable Cox proportional hazards model for time to recovery to the remote preoperative physical activity baseline.**

| Predictor | Hazard Ratio (HR) | 95% CI | P value |
| --- | --- | --- | --- |
| Age (per year) | 0.96 | 0.94–0.99 | 0.007 |
| BMI (per kg/m²) | 0.99 | 0.96–1.02 | 0.56 |
| Male sex (reference: female) | 1.16 | 0.78–1.72 | 0.46 |
| Surgery type: TKA (reference: THA) | 0.97 | 0.63–1.51 | 0.90 |
| Education: no college (reference: college) | 1.25 | 0.57–2.71 | 0.58 |
| Education: some college (reference: college) | 1.19 | 0.76–1.88 | 0.45 |
| Employment: working (reference: not working) | 0.67 | 0.42–1.05 | 0.08 |
| Income: low/skip (reference: high) | 1.1 | 0.63–1.92 | 0.73 |
| Income: mid (reference: high) | 0.76 | 0.48–1.20 | 0.24 |
| Marital status: not married (reference: married) | 0.87 | 0.56–1.33 | 0.51 |
| Immediate preoperative baseline (per 1000 steps) | 1.49 | 1.31–1.70 | <0.001 |
| Remote preoperative baseline (per 1000 steps) | 0.69 | 0.60–0.79 | <0.001 |

Remote preoperative baseline was defined as the mean daily step count during weeks −104 to −55 before surgery. Immediate preoperative baseline was defined as the mean daily step count during weeks −4 to −1 before surgery. Hazard ratios (HRs) greater than 1 indicate a higher likelihood of achieving recovery in physical activity. The proportional hazards assumption was assessed using scaled Schoenfeld residuals and was not violated. Definitions of categorical variables are detailed in the footnote of Table 1.

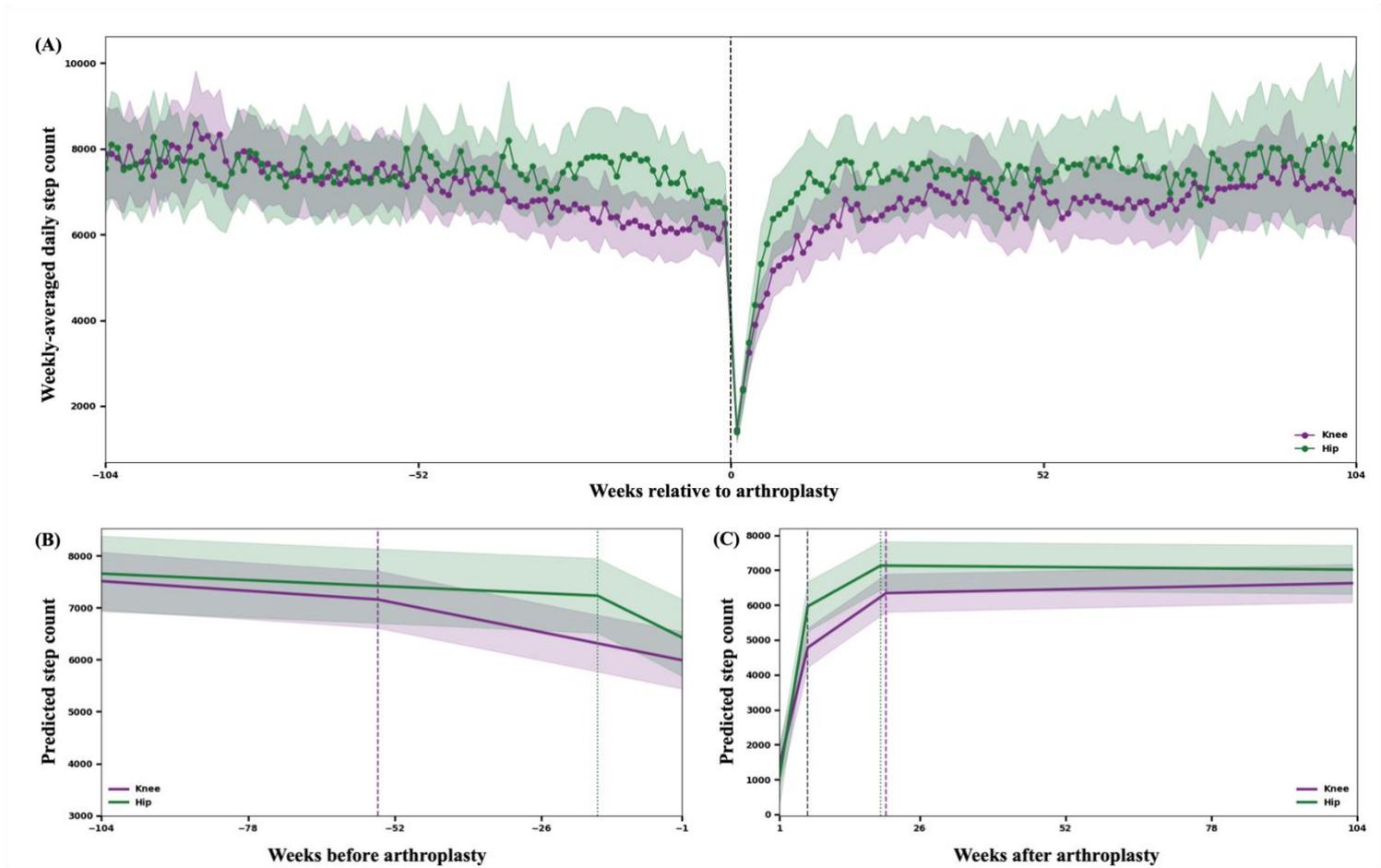

**Figure 1. Preoperative and Postoperative Physical Activity Trajectories Relative to Total Knee and Hip Arthroplasty.** (A) Observed weekly-averaged daily step counts relative to arthroplasty. (B) Piecewise model–estimated preoperative trajectory. (C) Piecewise model–estimated postoperative trajectory

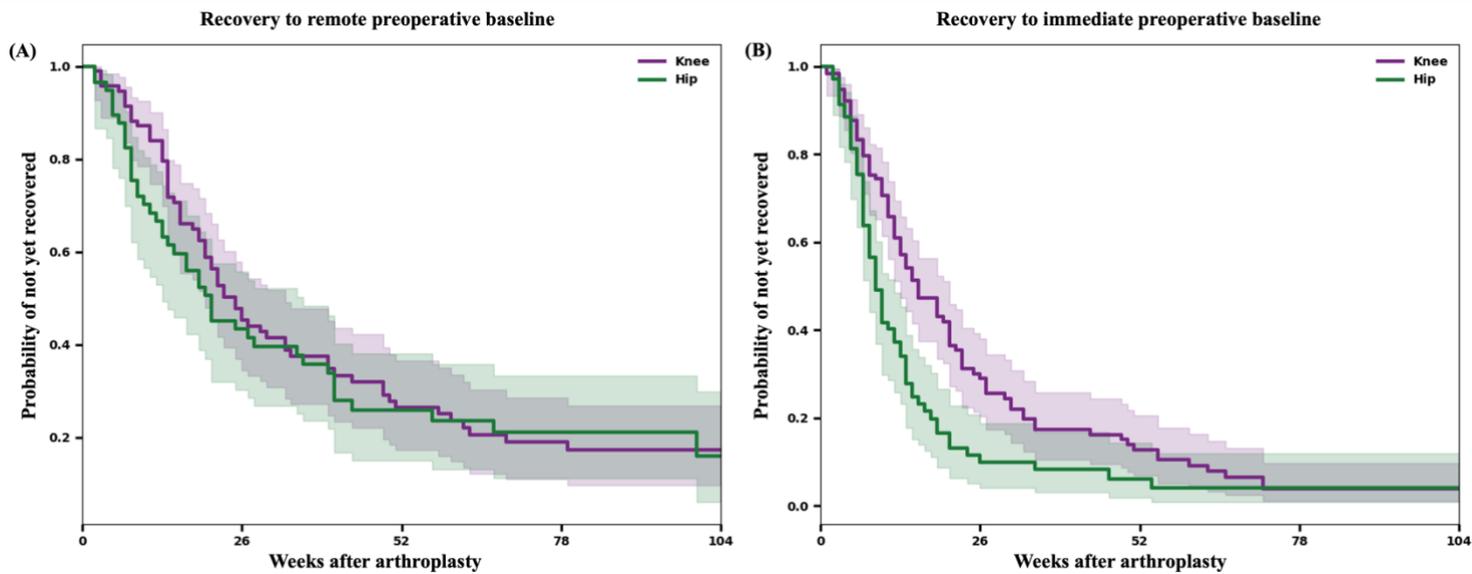

**Figure 2. Kaplan–Meier Estimates of Time to Physical Activity Recovery Following Total Knee and Hip Arthroplasty Under Different Preoperative Baseline Definitions.** (A) Recovery to remote preoperative baseline (mean daily step count during weeks −104 to −55 before arthroplasty). (B) Recovery to immediate preoperative baseline (mean daily step count during the 4 weeks preceding arthroplasty).

**Supplementary Table 1. Multivariable Cox proportional hazards model for time to recovery to Immediate preoperative baseline of physical activity.**

| Predictor | Hazard Ratio (HR) | 95% CI | P value |
|---|---|---|---|
| Age (per year) | 0.99 | 0.97–1.01 | 0.24 |
| BMI (per kg/m²) | 1.01 | 0.98–1.04 | 0.56 |
| Male sex (reference: female) | 0.96 | 0.65–1.41 | 0.83 |
| Surgery type: TKA (reference: THA) | 0.55 | 0.38–0.79 | 0.001 |
| Education: no college (reference: college) | 0.99 | 0.63–1.56 | 0.97 |
| Education: some college (reference: college) | 1.05 | 0.66–1.67 | 0.84 |
| Employment: working (reference: not working) | 0.8 | 0.54–1.18 | 0.27 |
| Income: low/skip (reference: high) | 1.37 | 0.85–2.20 | 0.19 |
| Income: mid (reference: high) | 1.39 | 0.91–2.11 | 0.13 |
| Marital status: not married (reference: married) | 0.92 | 0.63–1.34 | 0.66 |
| Immediate preoperative baseline (per 1000 steps) | 0.9 | 0.80–1.01 | 0.08 |
| Remote preoperative baseline (per 1000 steps) | 1.08 | 0.97–1.20 | 0.16 |

Immediate preoperative baseline was defined as the mean daily step count during the month preceding surgery. Remote preoperative baseline was defined as the mean daily step count during weeks −104 to −55 prior to surgery. HR >1 indicates faster recovery. The proportional hazards assumption was not violated. Definitions of categorical variables are detailed in the footnote of Table 1.